\documentclass   [a4paper,12pt]{article}
\usepackage{graphicx}
\begin{document}


\begin{center}
  {\bf  Controversy between Einstein and Bohr and two erroneous
                 arguments used in supporting Copenhagen quantum mechanics } \\ [3mm]
    Milo\v{s} V. Lokaj\'{\i}\v{c}ek \\
     Institute of Physics, AVCR, 18221 Prague 8, Czech Republic \\
\end{center}

Abstract

The support of Copenhagen quantum mechanics in the discussion
concerning EPR experiments has been based fundamentally on two
mistakes. First, quantum mechanics as well as hidden-variable
theory give the same predictions; the statement of Belinfante from
1973 about the significant difference must be denoted as mistake.
Secondly, the experimental violation of Bell's inequalities has
been erroneously interpreted as excluding the hidden-variable
alternative, while they have been based on assumption
corresponding to classical physics. The EPR experiments cannot
bring, therefore, any decision in the controversy between Einstein
and Bohr. However, the view of Einstein is strongly supported by
experimental results concerning the light transmission through
three polarizers.
 \\     [3mm]

In the controversy that started by two papers \cite{ein,bohr} in
1935 Einstein's view was refused by the then physical community.
The main reason consisted in the fact that any alternative of
hidden-variable theory was refused by von Neumann \cite{neu}. The
situation changed when Bell \cite{bell} showed that the refusal
was based on classical-physics assumption used by von Neumann.
Bell argued that the hidden-variable alternative was fully
admissible and derived also his famous inequalities. According to
Bell these inequalities  should have been fulfilled experimentally
for the hidden-variable theory, but violated in the case of the
quantum mechanics. However, they have been based on the
assumption, the impact of which has not been sufficiently
analyzed, as will be shown below.

Bell's results caused, however, that the original EPR Gedankenexperiment was
somewhat modified to be experimentally feasible. And the
coincidence polarization experiments started to be performed
practically in 1971. Decisive importance was given to them by
Belinfante \cite{belf} when he argued that the two competitive
theoretical alternatives (quantum mechanics and hidden-variable
theory) had to lead to mutually different predictions. However,
that has not been true as it will be shown now.

Polarized light going through two polarizers fulfills Malus law
\begin{equation}
    M(\alpha)\,=\,(1-\varepsilon)\cos^2(\alpha)\,+\,\varepsilon
    \label{mal}
\end{equation}
where $\alpha$ is the mutual deviation of polarizer axes;
expression (\ref{mal}) being valid for a pair of polarizers in
one-sided as well as coincidence arrangements. The standard
quantum mechanics deals, of course, with the so called ideal
polarizers only, where $\varepsilon=0$. For real polarizers it
holds always $\varepsilon>0$, for which the quantum mechanics has
phenomenological explanation only.

\begin{figure}[htb]
\begin{center}
\includegraphics*[scale=.35, angle= -90]{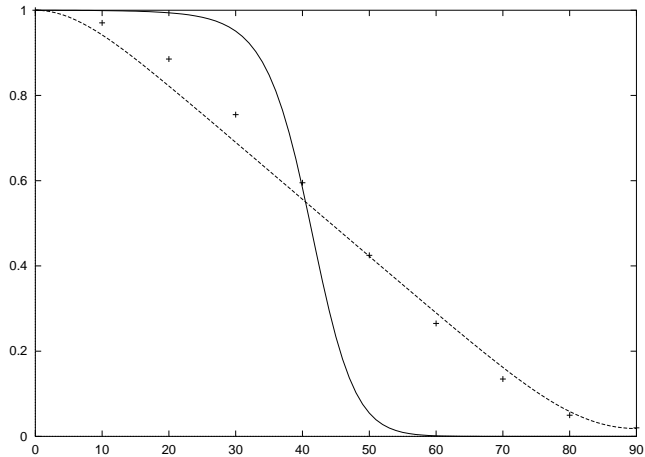}
\vspace{-2mm}
\caption   { {\it     Malus law and hidden variables;  $p_1(\lambda)$ - full line,
     $M(\alpha)$ calculated according to Eq. (\ref{palf}) - dashed line;
     actual Malus law (see Eq. (\ref{mal}) - $\varepsilon=0$)  - individual points. }}
\end{center}
 \end{figure}

In the hidden-variable alternative it is possible to write
\begin{equation}
 M(\alpha) \;=\;\int^{\pi/2}_{-\pi/2}d\lambda\;p_1(\lambda)\;p_1(\alpha-\lambda)
                         \label{palf}
\end{equation}
where $\lambda$ is Bell's hidden parameter (deviation of photon
polarization from the axis of the first polarizer). The function
$p_1(\lambda)$ represents the distribution of transmission
probability of non-polarized light through one polarizer and must
be found by fitting to obtain the Malus law according to Eq.
(\ref{palf}), which is well defined and soluble problem. A simple
approximate solution may be easily obtained if one puts, e.g.
     \[ p_1(\lambda) \;=\; 1- \frac{1-\exp(-(a|\lambda|)^{e})}{1 +
   c\exp(-(a|\lambda|)^{e})},  \;\;\;\;a=1.95, \;\;e=3.56, \;\;c=500;    \]
$p_1(\lambda) \;(=p_1(-\lambda))$ is represented by full line and
$M(\alpha)$ calculated for the given $p_1(\lambda)$ according to
Eq. (\ref{palf}) by dashed line in Fig. 1; $\lambda$ or $\alpha$
being shown on abscissa. The exact Malus law is represented by
individual points; see also \cite{lokj}
 \footnote {The quoted paper was submitted also to Physical Review,
but the publishing was refused by editorial board, which occurred
also earlier in the case of other papers of ours containing some
 critical points towards the standard quantum mechanics.}.
A much better agreement with the Malus law may be obtained with a
more flexibly parameterized function $p_1(\lambda)$.

To obtain the fundamentally different results Belinfante put quite
arbitrarily
             \[ p_1(\lambda)\,=\,\cos^2(\lambda) , \]
which differs significantly from the curve obtained by fitting and
shown in Fig.1; $\cos^2(\lambda)$ being represented e.g. by
individual points in Fig. 1. It should not be any surprise when
the EPR experiments (the first series having been finished in
1982, see \cite{asp}) have corresponded to the Malus law.
 And consequently, the polarization EPR experiments do not seem to be
suitable for bringing any decision between the quantum mechanics
and hidden-variable alternative.

However, an equally important problem has concerned the violation
of Bell inequalities as they have been interpreted mistakenly.
Until now nobody has analyzed the actual impact of the assumption
involved in their derivation. Their actual impact was studied and
shown for the first time in \cite{lok2}. The whole problem has
been then explained in a greater detail recently in the language
of the Bell operator (see \cite{lokj}).

\begin{figure}[htb]
\begin{center}
\includegraphics*[scale=.32, angle=-90]{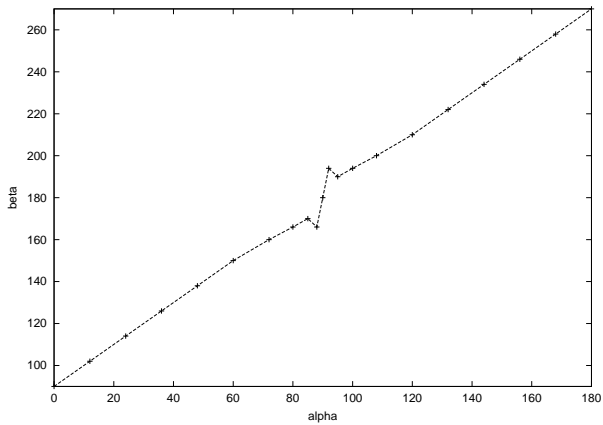}
  \caption { \it {  The angle $\,\beta\,$   corresponding to the minimum light
         transfer through three polarizers at a given $\,\alpha$.  } }
 \end{center}
 \end{figure}
\begin{figure}[htb]
\begin{center}
\includegraphics*[scale=.32, angle=-90]{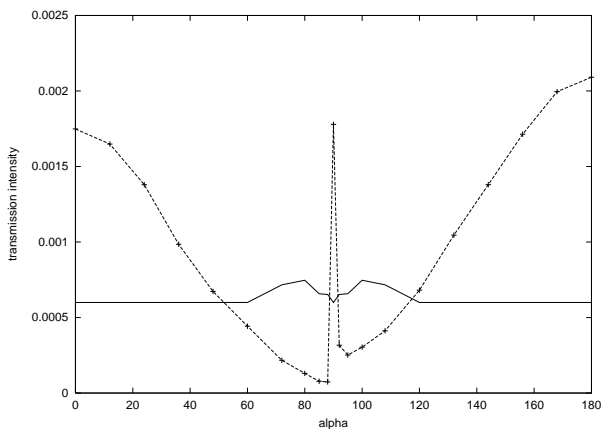}
  \caption { \it {    The measured transfer intensity through
  three polarizers (points on dashed line)
    corresponding to different pairs of angles $\,\alpha\,$  and $\,\beta\,$
    (taken from Fig. 1).  Full line: quantum-mechanical prediction
       (transmission through three ideal polarizers + additional
        phenomenological constant).  } }
 \end{center}
 \end{figure}

The consequences will be summarized here very shortly only. The
Bell operator is defined as
\begin{equation}
    B\;=\;a_1b_1+a_1b_2+a_2b_1-a_2b_2
 \end{equation}
where operators $a_j$ and $b_k$ represent measurements in
individual polarizers (being in coincidence arrangement). It holds
for expectation values of theses operators:
    \[ 0\; \leq |\langle a_j\rangle|, \; |\langle b_k\rangle| \leq\; 1 .   \]
The Bell operator may then have three different limit values   in
the dependence on the commutation relations between the operators
$a_j$ and $b_k$. It may hold (see \cite{lokj})
      \[ \langle B \rangle \;\leq\;\; 2,\;2\sqrt{2},\; 2\sqrt{3}  \]
where the individual limits correspond to: classical physics (all
operators $a_j$ and $b_k$ commuting mutually), hidden-variable
alternative, and Copenhagen quantum mechanics (no pair of
commuting operators).

To obtain his inequalities Bell had to use an assumption the
impact of which has not been sufficiently analyzed. It has been
assumed for it to correspond to hidden-variable alternative, while
in fact it has corresponded to classical physics (being
practically equivalent to that used earlier by von Neumann).
Therefore, it is only the classical-physics alternative that has
been excluded by the results of EPR experiments. And the decision
concerning the Einstein - Bohr controversy must be based on other
experiments.

Such experiments were performed already more than 10 years ago
when the transmission of light through three polarizers was
measured (see \cite{kra2}):
      \[       o----|---|^{\alpha}---|^{\beta}---> ;   \]
$\alpha$ and $\beta$ being deviations of the second and third
polarizers from the axis of the first polarizer. The main results
may be seen in Figs. 2 and 3.

Individual points in Fig. 3 represent experimental transmission
values in dependence on the angle $\,\alpha\,$ of the second
polarizer, which have been determined in the following way: In
individual cases the angle $\,\alpha\,$ was fixed and angle
$\,\beta\,$ of the third polarizer was established, so as the
light transmission through the whole triple was minimum
(corresponding $\beta$ shown always in Fig. 2).

Full line in Fig. 3 represents then quantum-mechanical prediction
for given angle pairs, having been estimated as the corresponding
dependence for the triple of ideal polarizers with an added
(unpredictable) constant. There is certainly a significant
disagreement between quantum-mechanical prediction and
experimental data; more details will be presented in \cite{lokm}.
\\
 \hspace*{5mm} And it is possible to summarize: \\
 - there have been two mistakes in arguments, on which the support
 of Copenhagen quantum mechanics has been founded until now; \\
 - in contradistinction to common opinion the results  of
polarization EPR experiments cannot bring any solution of the
controversy between Einstein and Bohr, when only the classical
 alternative has been excluded by these results;  \\
 - the remaining controversy (hidden-variables vs. Copenhagen
interpretation) may be decided e.g. on the basis of experiments
with three polarizers; decisive preference given to the view of
Einstein as quantum-mechanical prediction differs from
experimental results.

  \end{document}